\documentclass[twocolumn,amsfonts,showpacs,superscriptaddress,amsmath,amssymb,aps, longbibliography,prd]{revtex4-1}
\usepackage[normalem]{ulem}
\usepackage{dutchcal}
\usepackage{graphicx}% Include figure files
\usepackage{bm}% bold math
\usepackage{color}%\usepackage{hyperref}% add hypertext capabilities
\usepackage{alphabeta}
\usepackage[english]{babel}
\usepackage{dsfont}
\usepackage{xcolor}

\usepackage{epstopdf}
\usepackage{leftidx}
\newcommand{\bc}{\begin{center}}
\newcommand{\ec}{\end{center}}
\def\ba#1{\begin{array}{#1}\displaystyle}
\newcommand{\ea}{\end{array}}

\newcommand{\beq}{\begin{equation}}
\newcommand{\eeq}{\end{equation}}
\newcommand{\beqa}{\begin{eqnarray}}
\newcommand{\eeqa}{\end{eqnarray}}

\newcommand{\n}{\nonumber\\}
\newcommand{\bi}{\begin{itemize}}
\newcommand{\ei}{\end{itemize}}
\def\h#1{\hat{#1}}

\newcommand{\p}{\partial}

\newcommand{\bra}{\langle}
\newcommand{\ket}{\rangle}

\newcommand{\R}{{\mathbb{R}}}

\newcommand{\Tr}{{\rm Tr}}

\newcommand{\dd}{{\rm d}}

\usepackage[colorlinks, linkcolor=red, citecolor=blue]{hyperref}

\begin{document}

\title{Emergence of hydrodynamic spatial long-range correlations in nonequilibrium many-body systems}
 \author{Benjamin Doyon}
\affiliation
{Department of Mathematics, King’s College London, Strand, London WC2R 2LS, U.K.}
\author{Gabriele Perfetto}
\affiliation{Institut f\"ur Theoretische Physik, Eberhard Karls Universit\"at T\"ubingen, Auf der
Morgenstelle 14, 72076 T\"ubingen, Germany.}
  \author{Tomohiro Sasamoto}
\affiliation{Department of Physics, Tokyo Institute of Technology, Ookayama 2-12-1, Tokyo 152-8551, Japan}
\author{Takato Yoshimura}
\affiliation{All Souls College, Oxford OX1 4AL, U.K.}
\affiliation{Rudolf Peierls Centre for Theoretical Physics, University of Oxford,
1 Keble Road, Oxford OX1 3NP, U.K.}
%
%\date{\today}
%
\begin{abstract}
At large scales of space and time, the nonequilibrium dynamics of local observables in extensive many-body systems is well described by hydrodynamics. 
At the Euler scale, one assumes that each mesoscopic region independently reaches a state of maximal entropy under the constraints given by the available conservation laws.
%: locally, in each ``fluid cell", one finds a Gibbs or generalised Gibbs state.
Away from phase transitions, maximal entropy states show exponential correlation decay, and independence of fluid cells might be assumed to subsist 
during the course of time evolution. We show that this picture is incorrect: under ballistic scaling, regions separated by macroscopic distances {\em develop long-range correlations as time passes}. These correlations take a universal form that only depends on the Euler hydrodynamics of the model. They are rooted in the large-scale motion of interacting fluid modes, and are the dominant long-range correlations developing in time from long-wavelength, entropy-maximised states. They require {\em the presence of interaction} and {\em at least two different fluid modes}, and are of a fundamentally different nature from well-known long-range correlations occurring from diffusive spreading or from quasi-particle excitations produced in far-from-equilibrium quenches.
We provide a universal theoretical framework to exactly evaluate them, an adaptation of the macroscopic fluctuation theory to the Euler scale. 
We verify our exact predictions in the hard-rod gas, by comparing with numerical simulations and finding excellent agreement.
\end{abstract}

\maketitle

{\bf Introduction.}--- %The dynamics of many-body extended systems has been the focus of much recent theoretical and experimental research \cite{Bloch2008,Eisert_2015,Gogolin_2016,D_Alessio_2016,langen2016prethermalization}. 
Finding universal laws that govern many-body extended systems \cite{Bloch2008,Eisert_2015,Gogolin_2016,D_Alessio_2016,langen2016prethermalization} at large scales away from equilibrium is a fundamental problem in physics. Hydrodynamics is arguably the most far-reaching and successful such set of laws \cite{Spohn1991}. %Crucially, recent research has gone much beyond the conventional hydrodynamic equations of Navier-Stokes, applying the hydrodynamic principles to a large variety of systems.
The largest scale of hydrodynamics, the Euler scale, exists and is nontrivial as soon as the system admits ballistic transport  and interactions are on short enough distances; in particular, the system must possess at least a few extensive conserved quantities, and hydrodynamic modes are in one-to-one correspondence with these. Euler hydrodynamics applies to a wide array of many-body systems, including gases and fluids of interacting particles. A prominent example, which came to the fore recently, is the theory of generalised hydrodynamics (GHD) \cite{PhysRevX.6.041065,PhysRevLett.117.207201,PhysRevLett.121.160603,doyon_note_2017,PhysRevLett.120.045301,Bastianello_2022}, which proposes a universal structure for the Euler hydrodynamics of many-body integrable systems, and which has been shown to correctly describe cold atomic gases constrained to one dimension \cite{PhysRevLett.122.090601,PhysRevLett.126.090602,doi:10.1126/science.abf0147} and experimentally accessible gases of solitons \cite{El_2021,https://doi.org/10.48550/arxiv.2203.08551}. The Euler hydrodynamics of a microscopic model only requires the knowledge of basic aspects of the emergent degrees of freedom, such as their fluid velocities and their static correlations in stationary states. From these data, it makes a range of nontrivial physical predictions, including the large-scale motion of local observables, %when the system is initiated in a state with long-wavelength variations, %, in equilibrium states,
correlations at large separations in space-time \cite{SciPostPhys.3.6.039,10.21468/SciPostPhys.5.5.054,10.21468/SciPostPhysCore.3.2.016}, and the large-deviation theory for long-time ballistic transport \cite{2019,10.21468/SciPostPhys.8.1.007,10.21468/SciPostPhys.10.5.116,PhysRevE.102.042128}. 

Euler hydrodynamics is based on a simple extension of equilibrium thermodynamics \cite{Spohn1991,doyon_lecture_2019}: in every mesoscopic region, or ``fluid cell", the many-body system is assumed to maximise its entropy, under the local constraints provided by the extensive conserved charges $\hat{Q}_i = \int \dd x\,\hat{q}_i(x,t)$. Here $\hat{q}_i(x,t)$'s are the microscopic densities, related to the currents $\hat{j}_i(x,t)$ via continuity equations $\p_t \h q_i + \p_x \h j_i = 0$; we concentrate on one-dimensional systems for simplicity. Thus, in every fluid cell, a different Gibbs state, or generalised Gibbs ensemble (in integrable models) \cite{Vidmar_2016}, is reached: $\langle \bullet\rangle_{\underline\beta}=\Tr[ \mbox{exp}(-\sum_i \beta^i \hat{Q}_i) \bullet]/Z$ characterized by Lagrange parameters  $\underline{\beta}$. ``Mesoscopic" refers to a size, $L$, which is much greater than microscopic sizes $\ell_{\rm micro}$ -- such as inter-particle distances and interaction ranges -- but much smaller than the size $\ell$ where finite variations of local observables can be seen, $\ell_{\rm micro}\ll L \ll \ell$. Perhaps the most natural setup where these principles apply is when an initial state is prepared in the presence of external, long-wavelength fields. Maximising entropy with fields $\beta^i(x/\ell)$ coupled to the conserved densities and varying on large scales $\ell$, the state is
\beq
\label{init}
	\bra\bullet \ket_\ell = \Tr\Big(
	\mbox{exp}\Big[-\sum_{i}\int \dd x\,\beta^i_{\rm ini}(x/\ell) \hat{q}_i(x)\Big]\bullet
	\Big)/Z.
\eeq
The emergent, slow dynamics from the initial state \eqref{init} is that induced by the continuity equations for the mesoscopic densities $\mathtt q_i(x,t) = \bra \hat{q}_i\ket_{\underline \beta(x,t)}$ and currents $\mathtt j_i(x,t) = \bra \hat{\jmath}_i\ket_{\underline \beta(x,t)}$ as measured in the fluid cells,
\begin{equation}
\label{eq:eulerintro}
 \p_t \mathtt q_i(x,t) + \mathsf A_i^{~j}(x,t)\p_x \mathtt q_j(x,t) = 0,
\end{equation}
with $\mathsf A_i^{~j} = \p \mathtt j_i/\p \mathtt q_j$ the flux Jacobian. The mesoscopic densities are related to the generalized inverse temperatures $\underline{\mathtt{q}}\leftrightarrow \underline{\beta}$ bijectively thanks to positivity of 
the static covariance matrix $\mathsf C_{ij}=-\p\mathtt{q}_i/\p\beta^j$.

 Euler hydrodynamics asserts that the slow variation in space of the state \eqref{init} induces a corresponding slow variation in time, such that the state keeps the local equilibrium form; one may propose that at all times \eqref{init} correctly describes the states, with $\beta^i_{\rm ini}(x) \to \beta^i(x,t)$ (see, e.g., \cite{deGrootMazurNoneqThermo62,Spohn1991,doyon_lecture_2019}). In \eqref{init}, correlations vanish exponentially with the distance, under very broad conditions including non-zero temperature (finite $\beta^i$'s) and the lack of phase transition \cite{IsraelConvexity,BratelliRobinson2} (which we assume here). Thus, spatial correlations should vanish exponentially, even during 
time evolution; indeed each fluid cell is at ``local equilibrium", and entropy maximisation should occur independently in every fluid cell \cite{10.21468/SciPostPhys.5.5.054,10.21468/SciPostPhys.10.5.116,SciPCFT_breathing,bouchoule2022generalized}.

In fact, certain long-range, algebraic correlations are known to emerge in non-equilibrium situations when conservation laws are present. This is well studied for diffusive systems in non-equilibrium steady states (NESS) \cite{H_Spohn_1983, doi:10.1146/annurev.pc.45.100194.001241,OrtizdeZrate2004}: unbalanced thermostats at the system's boundaries lead to nonzero gradients, and correlations between conserved densities at macroscopic distances decay as (system size)$^{-1}$. This is due to the breaking of detailed balance at the diffusive scale and determined by viscous coefficients, and may be quantitatively described by fluctuating hydrodynamics and macroscopic fluctuation theory (see, e.g., \cite{doi:10.1146/annurev.pc.45.100194.001241,OrtizdeZrate2004,bertinilongrange}). But what happens at the Euler scale, where viscous effects are scaled down to zero size? 

In NESS emerging from the partitioning protocol in systems of infinite size \cite{Bernard_2016}, gradients vanish and correlations are weaker. The strongest are found in integrable systems, including free particles, where conserved density correlations decay as (distance)$^{-2}$ because of discontinuities in the occupation function of hydrodynamic modes \cite{PhysRevLett.120.217206,2035fbac601e4de9b4634047c14de12a}. But this decay is too quick to correlate Euler-scale fluid cells (see below). 

We note that a similar situation occurs at zero temperature, under the different physics of quantum fluctuations at Fermi points, and that a theory for the transport of such weak algebraic correlations on top of moving fluids is proposed in~\cite{SciPCFT_breathing,PhysRevLett.124.140603} (in GHD). Very far from equilibrium, stronger long-range correlations may develop: for instance, global quantum quenches generate finite densities of entangled particles that may propagate (diffusively or ballistically) and carry nontrivial entanglement \cite{Calabrese_2005,alba2017entanglement,alba2018entanglementlong} and correlations \cite{PhysRevA.89.053608,https://doi.org/10.48550/arxiv.2301.02326}. But entangled particle production is not expected to occur in long-wavelength states, Eq.~\eqref{init}.

Up to now, there has been no prediction, observation or theory for eventual long-range correlations emerging under ballistic scaling from \eqref{init}. The assumption of uncorrelated Euler-scale fluid cells, and that the form \eqref{init} stays valid in time, has remained, and appears to play an important role in recent studies of the evolution of correlations and fluctuations under inhomogeneous conditions and nonlinear hydrodynamic response theory \cite{10.21468/SciPostPhys.5.5.054,10.21468/SciPostPhys.10.5.116,SciPCFT_breathing,PhysRevLett.124.140603,Fava2021}.

In this manuscript, we show that {\em the assumption of uncorrelated Euler-scale fluid cells is generically incorrect}. We show that correlations %on macroscopic distances $\propto\ell$, which decay as $\ell^{-1}$,  
of strength $\propto\ell^{-1}$ develop dynamically from \eqref{init}, at macroscopic ($\propto \ell$) times and distances, under generic conditions for systems exhibiting ballistic transport. In particular, if $Q_i^R(\ell t),\, Q_j^{R'}(\ell t)$ are total charges lying on finite but macroscopically large regions $R,\,R'$ that are separated by a macroscopic distance, $|R|,|R'|\propto \ell,\,{\rm dist}(R,R')\propto \ell$, evaluated at macroscopic time $\ell t$, then their covariance is large, $\bra Q_i^R(\ell t) Q_j^{R'}(\ell t)\ket^{\rm c} \propto \ell$. This shows strong correlations between separated cells. The appearance of ballistically scaled long-range correlations at all macroscopic times is a general phenomenon which, to our knowledge, has not been discussed before. It holds no matter the nature of the system, quantum or classical, integrable or not, and is solely controlled by its Euler hydrodynamics.

This phenomenon is not explained by the theories for diffusive long-range correlations recalled above, as it does not depend on viscous coefficients or phenomenological noise, and occurs in ballistic times $t\propto x$. It gives the dominant correlations on large distances, beyond diffusive broadening and of larger strength than the $1/x^2$ correlations due to occupation discontinuities. It is not due to quasi-particle excitations, as it is a universal hydrodynamic effect. By contrast, we show that the phenomenon occurs in long-wavelength inhomogeneous state (as in \eqref{init}), only if the Euler hydrodynamic theory is interacting, and only if it admits at least two different fluid modes (with different velocities).
Euler-scale long-range correlations invalidate the assumption that on every time-slice a state such as \eqref{init} is found. This thus calls for a new understanding of the principles of Euler hydrodynamics, and a re-think of recent studies of hydrodynamic nonlinear response and the evolution of correlations and fluctuations.

We quantify this phenomenon by proposing that the principle replacing independent local entropy maximisation of fluid cells is that of {\em relaxation of fluctuations}: local observables relax to fixed, non-fluctuating functions of conserved densities, which themselves fluctuate. This is developed into a universal theory, the ballistic macroscopic fluctuation theory (BMFT). The BMFT is a hydrodynamic large-deviation theory, solely based on the emergent Euler hydrodynamic data of the model, which characterises all fluctuations and correlations at the ballistic hydrodynamic scale, including under fluid motion.

For illustration, we study the paradigmatic hard-rod model of statistical physics, which is simple enough to be amenable to high-accuracy numerical simulations, yet truly interacting. We find that the model does indeed develop long-range correlations, which are quantitatively in excellent agreement with our theory.

{\bf Ballistic long-range correlations.}--- We show that correlations in the initial state \eqref{init}, between macroscopically separated observables $\h o_1(\ell x_1,\ell t)$ and $\h o_2(\ell x_2, \ell t)$ at macroscopic times, generically has strength $\ell^{-1}$. That is, the connected correlation function $\bra \h o_1 \h o_2\ket^{\rm c} := \bra \h o_1 \h o_2\ket - \bra \h o_1\ket\bra \h o_2\ket$ has a {\em nonzero Euler-scaling limit}
\footnote{Technically, for Eq.~\eqref{S} to be generically true, on its right-hand side observables $o_i(\ell x_i,\ell t)$ should be replaced by their fluid-cell means by averaging on cells in space-time of size $L$ around $\ell x_i,\ell t$.}
:
\beq
\label{S}
	S_{\h o_1,\h o_2}(x_1,t;x_2,t):=\lim_{\ell\to\infty} \ell \bra \h o_1(\ell x_1,\ell t) \h o_2(\ell x_2,\ell t)\ket^{\rm c}_\ell \neq 0,
\eeq
when $x_1\neq x_2$. If fluid cells were uncorrelated, the result would be a delta function. We will demonstrate that 
there is an additional nonzero contribution if three conditions are satisfied: (\hypertarget{condition_inh}{{\color{red} a}}) the initial state is inhomogeneous (so there is fluid motion); (\hypertarget{condition_int}{{\color{red} b}}) the system is interacting (its hydrodynamic flux Jacobian $\mathsf A_i^{~j}$ depends on the state); and (\hypertarget{condition3_twomodes}{{\color{red} c}}) the model has at least two hydrodynamic modes with different fluid velocities ($\mathsf A_i^{~j}\not\propto \delta_i^{~j}$). The latter two conditions are generic in many-body systems that admit ballistic transport. %-- that is, with nontrivial Euler hydrodynamics. 
This includes translation invariant Hamiltonian systems such as the anharmonic chain \cite{Spohn2014}, integrable quantum spin chains \cite{PhysRevLett.117.207201}, and integrable gases such as the one-dimensional interacting Bose gas \cite{PhysRev.130.1605} and soliton gases \cite{El_2021}. Taking $\h o_1 = \h q_i,\,\h o_2 = \h q_j$ and integrating over the regions $R,R'$, the result implies that $\bra Q_i^R(\ell t) Q_j^{R'}(\ell t)\ket^{\rm c} \propto \ell$ as claimed above.

The result \eqref{S} means that strong algebraic correlations develop at large times and distances. If the initial state has short-range correlation, which is the case for \eqref{init}, then as time passes, correlations grow, and at macroscopic times (of the order of the variation scale of the initial state), they are present on linearly growing macroscopic distances (set by the maximal fluid velocity), and are of strength inversely proportional to the macroscopic scale. Their shape on the macroscopic scale and the way they eventually decay in time, i.e., the precise function $S_{\h o_1,\h o_2}(x_1,t;x_2,t)$ -- depends on the model and the precise shape of the initial condition.
%As this is on macroscopic distances and times, it means that macroscopically separated points have correlations that decay more slowly than exponentially -- thus fluid cells develop in time mutual correlations.

The physics of such correlations comes from nonlinear hydrodynamic response. \textit{Correlated fluid modes} are emitted from fluid cells in response, at nonlinear order, to the change of the cells' states in time.
%Due to a nonlinear hydrodynamic response of the system, there exist \textit{“coherent" fluid modes}, which are emitted in pairs or larger groups with 
%correlations, from fluid cells which move from their initial positions due to inhomogeneity. 
The long-range correlations are the product of such 
coherent fluid modes of different velocities travelling to macroscopic distances and scattering amongst each other. Nonlinear response is possible only if interaction is present, and long-range correlations can only occur, after the emission event, if fluid modes propagate at different velocities; this explains the three conditions above. This is a new mechanism, inherently nonlinear and distinct from the well-known mechanism for time correlations due to hydrodynamic linear response \cite{10.21468/SciPostPhys.5.5.054,10.21468/SciPostPhysCore.3.2.016,10.21468/SciPostPhys.10.5.116,Fava2021}, whereby Euler-scale correlations occur when a mode emitted by $\h o_1$  at time $t_1$ is probed by $\h o_2$ at $t_2>t_1$. The emission of coherent fluid modes parallels, but is different from, the mechanism of entangled particle production found to give long-range correlations in homogeneous quantum quenches both for diffusive \cite{PhysRevA.89.053608} and ballistic \cite{https://doi.org/10.48550/arxiv.2301.02326} transport. Naturally, the latter is not universal and does not require inhomogeneity or interaction (in the above hydrodynamic sense), by contrast to the effect we have uncovered.

In order to prove and quantify the result, Eq.~\eqref{S}, we introduce the ballistic macroscopic fluctuation theory. %The BMFT is an hydrodynamical large-deviation theory, solely based on the emergent Euler hydrodynamics data of the model, that describes fluctuations and correlations at the Euler-ballistic hydrodynamic scale. 
The BMFT, which applies to the ballistic scale, is similar in spirit macroscopic fluctuation theory (MFT) \cite{2002,RevModPhys.87.593,2006,2009,PhysRevLett.113.078101,2015,mms,Derrida_2007}, for the diffusive scale. Technically, the long-range correlations that occur at ballistic and diffusive scales have similar explanations in BMFT and MFT, respectively: the non-locality of a ``free-energy functional" emerging after time evolution. The physics is nevertheless markedly different; the former is not an effect of noise, and requires the presence of at least two fluid modes, in contrast to the latter. For the full development of the BMFT, see the companion manuscript \cite{https://doi.org/10.48550/arxiv.2206.14167}.

\begin{figure}[h!]
\centering
\includegraphics[width=\linewidth]{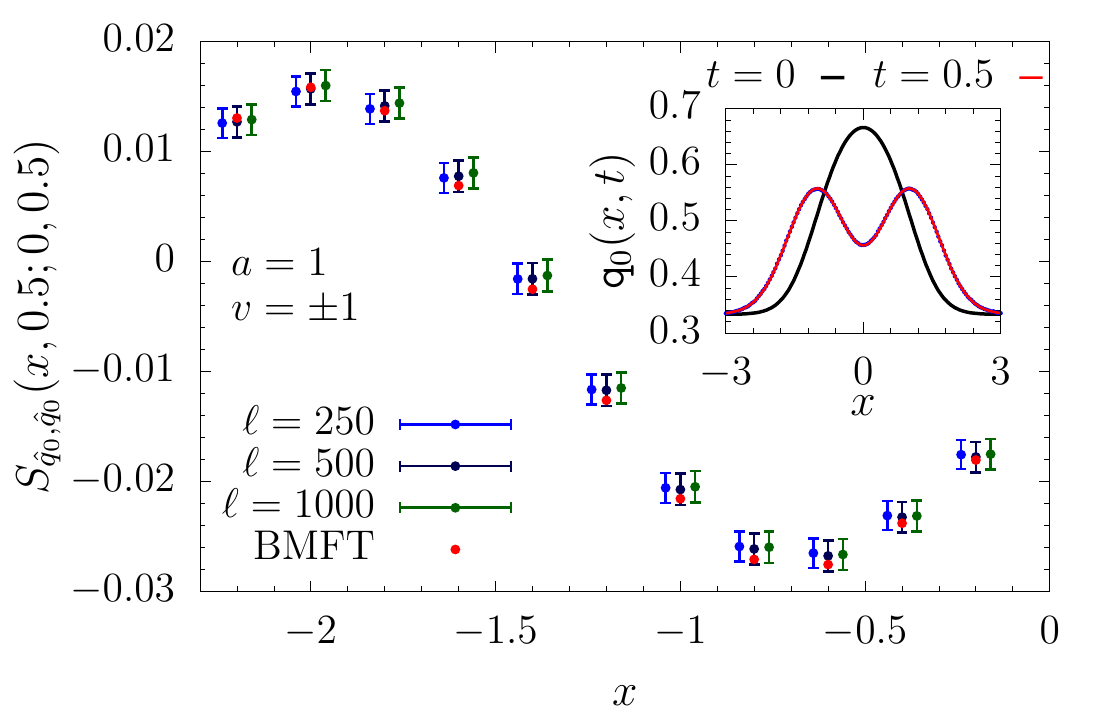}
\caption{{\bf Bump release of the hard-rod gas.} Rod density mean value $\mathtt{q}_0(x,t)$ (in the inset) and equal-time connected correlation function $S_{\h q_0,\h q_0}(x,t;0,t)$ from \eqref{S} with $\h q_0(x) = \sum_i \delta(x-x_i)$,
%, in Eq.~\eqref{eq:connected_correlator_fluid_cell_hr}, 
evaluated at the macroscopic time $t=0.5$ as a function of the macroscopic space coordinate $x$. The initial state is the density bump in Eq.~\eqref{eq:initial_state_bump} (black solid line in the inset). Three values $\ell=250,500,$ and $1000$ are reported with the associated uncertainty bars (the data for the scales $\ell=250$ and $1000$ are shifted horizontally by $\pm 0.04$ to make them clearly visible). The collapse of the numerical data is evident, numerically confirming the existence of Euler-scale long-range correlations of order $10^{-2}$. These correlations are captured by the BMFT (red points) with an excellent agreement.
%The numerical data, corresponding to three different macroscopic scales $\ell=250,500$ and $1000$, are reported with the associated uncertainty bars. 
%The rod length is $a=1$, while velocities can take only two values $v=\pm 1$ with equal probability.
The numerical data are obtained by simulating the deterministic hard-rod dynamics and by averaging over a large number of independent realizations of the rods initial positions and velocities.  In the inset, the numerical data (blue dotted curve) are obtained with $\ell=250$ and match the solution of the Euler equation \eqref{eq:eulerintro} (red solid line).} \label{fig:bump_cen_corr_intro}
\end{figure}

{\bf Example: the hard-rod gas.}--- Before developing the BMFT, we illustrate the result on the hard-rod model, with a well-established Euler hydrodynamics description \cite{Boldrighini1983,boldrighini1997,Doyon_2017,SciPostPhys.3.6.039}. The rods with unit length $\ell_{\rm micro} = a=1$ have ordered positions $x_i(t)\in\R$ that move in time $t$ freely at velocities $v_i$, except at collisions $x_{i+1}(t)-x_{i}(t) = 1$ where velocities are exchanged. We consider the rod density $\h q_0(x) = \sum_i \delta(x-x_i)$ and evaluate $S_{\h q_0,\h q_0}(x,t;0,t)$ from \eqref{S}. %the following object:
%\beq
%    S_{\h q_0,\h q_0}(x,t;0,t) =  \lim_{\ell\to\infty}  \ell \bra \h q_0(\ell x,\ell t)\h q_0(0,\ell t)\ket_\ell^{\rm c}.
%\label{eq:connected_correlator_fluid_cell_hr} 
%\eeq
We take large but finite values of $\ell$; we find that $\ell> 100$ is sufficient for a $<10\%$ relative accuracy. The initial state $\bra \cdots\ket_\ell$ is a random distribution of nonoverlapping rods, uniformly with velocities $v_i\in\{\pm1\}$ and with the ``bump'' density profile
\beq
	\bra \h q_0(x,0)\ket_\ell = \frac{1+3 e^{-(x/\ell)^2}}{3+3 e^{-(x/\ell)^2}}\in[1/3,2/3].
\label{eq:initial_state_bump}
\eeq
 The particle density $\mathtt{q}_0(x,t)$ is shown in the inset of Fig.~\ref{fig:bump_cen_corr_intro}. At the macroscopic time $t=0.5$ the initial bump splits into two counterpropagating bumps. 
 We observe that the correlation $\bra \h q_0(\ell x,\ell t) \h q_0(0,\ell t)\ket^{\rm c}_\ell$ decay as $1/\ell$ and find a finite result as $\ell\to\infty$ by multiplying by $\ell$, in accordance with the definition in Eq.~\eqref{S} of the Euler scaling limit. Note how both positive and negative correlations occur, and how one point, $x=0$, is not in the moving bump; these are not simply correlations between the counter-propagating bumps. The comparison between the result predicted  by BMFT and the simulation is displayed in Fig.~\ref{fig:bump_cen_corr_intro}. Striking are both the fact that the result is nonzero, thus long-range correlations are present, and that the BMFT correctly predicts it. Correlations develop everywhere on the region $x\in (-2,2)$, determined by the maximum fluid velocity, where the large-scale motion takes place. Outside this interval, the state at time $t=0.5$ still retains its homogeneous equilibrium structure and long-range correlations are zero.

{\bf BMFT.}--- As recalled, Euler-scale hydrodynamics is justified through local relaxation: local-observable averages are functions of the densities $\underline{\mathtt{q}}$ only, %, which allows us to write the local equilibrium average of an observable $\hat{o}$ as 
$\lim_{\ell\to\infty}\langle\hat{o}(\ell x,\ell t)\rangle_\ell=\langle\hat{o}\rangle_{\underline{\beta}(x,t)}:=\mathtt{o}[\underline{\mathtt{q}}(x,t)]$ \cite{Spohn1991,doyon_lecture_2019}. % where $\bra\cdot\ket_{\underline\beta(x,t)}$ is the (generalised) Gibbs state at the fluid cell of space-time coordinates $x,t$. 
Applying this assumption to the current averages $\mathtt{j_i}[\underline{\mathtt{q}}(x,t)]$, the Euler equation \eqref{eq:eulerintro} follows. 
The fundamental principle of the BMFT extends this  assumption to fluctuations. We assume that mesoscopic charge densities can be identified with classical fluctuating variables $\hat{\underline{q}}(\ell x,\ell t):=\underline{q}(x,t)$ \footnote{
For systems with multiple conservation laws, one has a vector $\underline{q}$ of fluctuating variables with components $q_i$. The symbols $\hat{q}_i$ and $\mathtt{q}_i$ represent the associated operator and its (generalized) Gibbs average, respectively.}. {\em Local relaxation of fluctuation} then states that mesoscopic fluctuating variables $\hat{o}(\ell x,\ell t):=o(x,t)$ do not fluctuate independently but are fixed functions of $\underline{q}(x,t)$. The functional dependence of $o(x,t)$ on $\underline{q}(x,t)$ is shown to be completely determined by the (generalized) Gibbs state: $o(x,t)=\mathtt o[\underline{q}(x,t)]$ \cite{https://doi.org/10.48550/arxiv.2206.14167}. Applying this to currents $j_i(x,t)=\mathtt j_i[\underline{q}(x,t)]$, and exploiting the conservation laws, the BMFT probability measure for the fluctuating charge densities is obtained,
\begin{equation}
    \dd \mathbb{P}[\underline{q}(\cdot,\cdot)] = \dd \mu [\underline{q}(\cdot,\cdot)]\,e^{-\ell \mathcal{F}[\underline{q}(\cdot,0)]}\delta[\p_t \underline{q}+\p_x \underline{\mathtt{j}}[\underline{q}]],
\label{eq:BMFT_measure}    
\end{equation}
where $\dd \mu [\underline{q}(\cdot,\cdot)]$ is the flat measure for functions on the space-time region $\mathbb{S}:=\mathbb{R}\times[0,T]$. This implies that fluctuations of mesoscopic densities stem from those of the initial state and are deterministically propagated in space-time according to the Euler equation. Initial fluctuations of densities $\underline{q}(\cdot,0)$ are determined by $e^{-\ell \mathcal{F}[\underline{q}(\cdot,0)]}$; the large deviation function for the state \eqref{init} is:
\begin{equation}
\label{eq:init_ldev}
\mathcal{F}[\underline{q}(\cdot,0)]
    \!=\!\int_\mathbb{R}\dd x\, \Big(\beta^i_\mathrm{ini}(x)q_i(x,0) \!-\!f[\underline{\beta_{\rm ini}}(x)]\!-\!s[\underline{q}(x,0)]\Big),
\end{equation}
with $s[\underline{q}]$ and $f[\underline{\beta}]$ the entropy and free-energy density, respectively.  Equation \eqref{eq:init_ldev} follows from statistical mechanics \cite{Derrida_2007}. It is also used in the MFT (for diffusive systems, where, however, the measure \eqref{eq:BMFT_measure} in space-time is different); see, e.g., Refs.~\cite{2009,PhysRevLett.113.078101}. One may consider a more general initial state, e.g., itself with long-range correlations \cite{Jin_2021}. The BMFT average is given by
\begin{equation}
     \langle\langle\bullet\rangle\rangle_\ell
    = \frac{1}{Z}\int_{(\mathbb S)}\dd\mu[\underline{q}(\cdot,\cdot)]\,
    e^{-\ell \mathcal F[\underline{q}(\cdot,0)]}
    \delta(\p_t \underline {q}+\p_x \underline {\mathtt{j}}[\underline{q}])
    \bullet,
\end{equation}
and reproduces the Euler scale of $\langle\bullet\rangle_\ell$ at large $\ell$. The delta function is represented via Laplace transform, leading to the introduction of an auxiliary field $\underline{H}(x,t)$.
%\begin{align}
%    &\delta[\p_t \underline{q}+\p_x \underline{\mathtt{j}}[\underline{q}]]=\n
%    & \int_{(\mathbb{S})}\dd\mu[\underline{H}(\cdot,\cdot)]\exp  \Bigl[-\ell\int_\mathbb{R}\dd x\int_0^T\dd t\, H^i(\p_tq_i+\p_x\mathtt{j}_i[\underline{q}])\Bigr]\!,
%\end{align}
%where $\underline{H}(x,t)$ is an auxiliary field. 
In the large-$\ell$ limit, all results are obtained by saddle-point calculations -- thus we have reduced the problem of Euler-scale correlations to a functional minimisation problem. From Eq.~\eqref{eq:BMFT_measure}, all BMFT predictions follow.

Note that both the BMFT and the MFT \cite{RevModPhys.87.593} are large-deviation theories based on an action formalism describing space-time configurations of densities and currents. They, however, pertain to {\em different hydrodynamic scales}, ballistic and diffusive, respectively, and as such they cannot be derived from one another.
In the companion manuscript \cite{https://doi.org/10.48550/arxiv.2206.14167}, we also combine these and introduce a multiscale hydrodynamic fluctuation theory describing both the ballistic and the diffusive scale.

{\bf Long-range correlations from the BMFT.}--- From the above we evaluate \eqref{S}, for conserved densities, using the BMFT measure \eqref{eq:BMFT_measure} as
\begin{equation}
S_{\hat{q}_{i_1},\hat{q}_{i_2}}(x_1,t_1;x_2,t_2)=\lim_{\ell\,\to\infty} \ell\langle\langle q_{i_1}(x_1,t_1)
q_{i_2}(x_2,t_2)\rangle \rangle^{\rm c}_\ell.
\label{eq:euler_correlator_BMFT}
\end{equation}
We introduce a generating function $\exp(\ell\Lambda)$ with $\Lambda=\lambda_1 q_{i_1}(x_1,t_1)+\lambda_2q_{i_2}(x_2,t_2)$. By saddle point, we have as $\ell\to\infty$, $\langle\langle\exp (\ell \Lambda)\rangle\rangle_\ell\to\exp(-\ell \mathcal F_\Lambda[\underline {q^*}])$ where $ \mathcal F_\Lambda[\underline q(\cdot,\cdot)] = \mathcal F[\underline {q}(\cdot,0)] - \Lambda[\underline {q}(\cdot,\cdot)]$. One calculates $ S_{\hat{q}_{i_1},\hat{q}_{i_2}}(x_1,t_1;x_2,t_2)$ by taking derivatives with respect to $\lambda$ at $\lambda=0$. Dropping the superscript $*$, $\underline q$ (no longer fluctuating) solves the saddle point equations:
\begin{subequations}
\label{mftcorrgeneric}
\begin{align}
     H^i(x,0)&=\beta^i_\mathrm{ini}(x)-\beta^i(x,0),\label{mftcorrgeneric1}\\
      H^i(x,T)&=0,\label{mftcorrgeneric2}
      \\ \label{mftcorrgeneric3}
     \p_t\beta^i+\mathsf A_j^{\,\,i}[\underline\beta]\p_x\beta^j&=0,\\
      \p_tH^i+\mathsf A_j^{\,\,i}[\underline\beta]\p_xH^j &=-\lambda\delta^{~i}_{i_1}\delta(x-x_1)\delta(t-t_1),  \label{mftcorrgeneric4}
\end{align}
\label{mftcorrgeneric4_set}%
\end{subequations}
with the boundary condition $\beta^i(x\to \pm \infty,t)=\beta^i_{\mathrm{ini}}(x \to \pm \infty)$ and $H^{i}(x\to\pm\infty,t)=0$.
Manipulations of these BMFT equations generically imply  long-range correlations, giving, in particular,
\begin{align}
\label{dynamcorr2}
     &S_{\hat{q}_{i_1},\hat{q}_{i_2}}(x_1,t_1;x_2,t_1)=\mathsf C_{i_1i_2}(x_1,t_1)\delta(x_2-x_1) \n
    &-\p_\lambda\left.\left(U_\lambda(t_1,0)\underline{\beta_\mathrm{ini}}\right)^{i}(x_2)\right|_{\lambda=0}\mathsf C_{i\,i_2}(x_2,t_1),
\end{align}
where $U_\lambda(t,t')$ is the nonlinear time-evolution operator associated with Eq.~\eqref{mftcorrgeneric3}, $\underline{\beta}(t)=U_\lambda(t,t')\underline{\beta}(t')$. The first term in \eqref{dynamcorr2} is the linear response contribution \cite{10.21468/SciPostPhys.5.5.054,10.21468/SciPostPhysCore.3.2.016,10.21468/SciPostPhys.10.5.116,Fava2021},
the correlation within the fluid cell, supported at $x_1=x_2$. The second term is generically nonzero for $x_1 \neq x_2$ (the operator $U_{\lambda}$ depends on $x_1$) and accounts for long-range correlations.
%It describes nonlinear correlated normal modes production and scattering through all the times prior to $t_1$.
It vanishes when $t_1=0$ [$U_{\lambda}(0,0)=1$], as the initial state \eqref{init} has no Euler-scale correlations; and also when the state is homogeneous [$U_{\lambda}(t_1,0)\underline{\beta_\mathrm{ini}}=\underline{\beta_\mathrm{ini}}$] and the model is noninteracting [$U_{\lambda}$ does not depend on $\lambda$],
%In such cases linear response theory applies.
giving conditions (\hyperlink{condition_inh}{a}) and (\hyperlink{condition_int}{b}) (see after Eq.~\eqref{S}). The condition (\hyperlink{condition3_twomodes}{c}) follows from hydrodynamic projections \cite{Spohn1991,Kipnis1999,Ayala2018,SciPostPhys.3.6.039,10.21468/SciPostPhys.5.5.054,SPOHN1982353,Doyon2022}, see \cite{https://doi.org/10.48550/arxiv.2206.14167}.

In general, it is challenging to bring the second term on the right-hand side of Eq.~\eqref{dynamcorr2} to a calculable form. Solving the BMFT equations \eqref{mftcorrgeneric1}-\eqref{mftcorrgeneric4} is generically tricky as shock solutions, where entropy is not conserved, may appear \cite{bressan2000hyperbolic,PhysRevLett.119.195301}. In integrable models, this problem does not arise, as GHD is known to display no shock solutions. In GHD, Eqs.~\eqref{mftcorrgeneric1}-\eqref{mftcorrgeneric4} can be exactly solved using the method of characteristics \cite{Doyon2018} leading to an expression for $S_{\hat{q}_{i_1},\hat{q}_{i_2}}(x_1,t_1;x_2,t_1)$
%=\mathsf C_{i_1i_2}(x_1,t_1)\delta(x_1-x_2)+E_{i_1i_2}(x_1,x_2;t_1)$ for \eqref{dynamcorr2}. The function $E_{i_1i_2}(x_1,x_2;t_1)$ is obtained by solving
in terms of solutions to certain integral equations, see \cite{https://doi.org/10.48550/arxiv.2206.14167}. In the case of the hard-rod model, this leads to the result reported in Fig.~\ref{fig:bump_cen_corr_intro}. 

%but for integrable systems, the machinery of GHD greatly facilitates the computation, allowing us to obtain the exact expression of it. We refrain from writing down the explicit result here as it is in general rather cumbersome even for hard-rods, but in \cite{https://doi.org/10.48550/arxiv.2206.14167}, we show that the Euler scale correlators take the form $S_{\hat{q}_{i_1},\hat{q}_{i_2}}(x_1,t_1;x_2,t_1)=\mathsf C_{i_1i_2}(x_1,t_1)\delta(x_1-x_2)+E_{i_1i_2}(x_1,x_2;t_1)$, where $E_{i_1i_2}(x_1,x_2;t_1)$ is given by thermodynamic quantities only.

{\bf Conclusions.}--- We have shown that long-range correlations \eqref{S} generically develop under ballistic scaling over time in many-body systems. Here (and the companion manuscript \cite{https://doi.org/10.48550/arxiv.2206.14167}), we have provided, to our knowledge, the first observation and description of this universal phenomenon; This happens under three generic conditions: inhomogeneity in initial conditions, interaction, and the presence of more than one fluid velocity. It is different from known long-range effects seen in diffusive NESS and in global quenches. This result invalidates the fundamental tenet of Euler hydrodynamics that at each time fluid cells undergo entropy maximization independent of each other. By introducing the BMFT, the first hydrodynamic fluctuation theory describing all correlations and fluctuations  
at the Euler scale, 
we have evaluated the ballistic long-range correlations in the hard-rod model observing excellent agreement with numerical simulations. The long-range correlations we unveil are expected to have significant impact and open new research directions in the field of correlations and fluctuations in inhomogeneous fluids, which is a broad area of current active research. 
For example, can we extend quantum GHD \cite{PhysRevLett.124.140603} to introduce such stronger correlations from nonlinear effects at zero or small temperatures? Do they affect nonlinear response coefficients \cite{Fava2021}? 
We also expect that these long-range correlations can be experimentally observed using quantum gas microscopes \cite{https://doi.org/10.48550/arxiv.2107.00038}.

{\bf Acknowledgements}---We are grateful to Bruno Bertini, Olalla Castro-Alvaredo, Jacopo De Nardis, Fabian Essler, Tony Jin, Pierre Le Doussal, Adam Nahum, Tibor Rakovszky, Paola Ruggiero, and Herbert Spohn for illuminating discussions. The work of B.D. was supported by the Engineering and Physical Sciences Research Council (EPSRC) under grants EP/W000458/1 and EP/W010194/1. G.P. acknowledges support from the Alexander von Humboldt Foundation through a Humboldt research fellowship for postdoctoral researchers.
The work of T.S. has been supported by JSPS KAKENHI Grant Nos. JP16H06338, JP18H03672, JP19K03665, JP21H04432, JP22H01143. B.D., T.S. and T.Y. acknowledge hospitality and support from the Galileo Galilei Institute, and from the scientific program on ``Randomness, Integrability, and Universality''. B.D. and T.Y. acknowledge hospitality and support from the Isaac Newton Institute, and from the program ``Dispersive hydrodynamics: mathematics, simulation and experiments, with applications in nonlinear waves".

\bibliographystyle{apsrev4-1}
\bibliography{bib.bib}

%\begin{thebibliography}{99}
%\bibitem{bla} Bla
%\end{thebibliography}%

\end{document}